\documentclass[aps,twocolumn, prd, showkeys, preprintnumbers, superscriptaddress, nobibnotes, noshowpacs, nofootinbib]{revtex4-1}

\usepackage{amsmath}
\usepackage{amsfonts}
\usepackage{amssymb}
\usepackage{bbold}
\usepackage{epsfig}
\usepackage{graphicx}
\usepackage{subfigure}
\usepackage{bm}
\usepackage{array}
\usepackage{mathtools}
\usepackage{listings}
\usepackage{color}
\usepackage[normalem]{ulem}
\usepackage[colorlinks=False,pdfusetitle]{hyperref}
\usepackage{lipsum}
\definecolor{Green}{RGB}{0, 128, 0}
\graphicspath{{./Figs/}}

\begin{document}

\title{Differential distributions for  Single  Top Quark Production at the LHeC}

\author{Meisen Gao}
\email{gmason@sjtu.edu.cn}
\affiliation{INPAC, Shanghai Key Laboratory for Particle Physics and Cosmology, School of Physics and Astronomy, Shanghai Jiao-Tong University, Shanghai 200240, China}

\author{Jun Gao}
\email{jung49@sjtu.edu.cn}
\affiliation{INPAC, Shanghai Key Laboratory for Particle Physics and Cosmology, School of Physics and Astronomy, Shanghai Jiao-Tong University, Shanghai 200240, China}
\affiliation{Key Laboratory for Particle Astrophysics and Cosmology (MOE), Shanghai 200240, China}
\affiliation{Center for High Energy Physics, Peking University, Beijing 100871, China}

\date{\today}

\begin{abstract}
{We present a phenomenological study of the single top (anti-)quark production with leptonic decays at the Large Hadron electron Collider (LHeC) at the next-to-leading-order (NLO) in QCD. We focus on various differential distributions in a fiducial region. The NLO corrections can reduce the fiducial cross section by 14\%.  We find the NLO predictions exhibit strong stability under scale variations for most observables considered while the scale variations at the leading-order (LO) dominated in the theoretical uncertainties. We propose a method of determining the top-quark mass using the measurement of the average transverse momentum of the charged lepton. The scale variations at the NLO induce a theoretical uncertainty of about 1.3 GeV of the extracted top-quark mass. The statistical error of the extracted top-quark mass amounts to 1.1 GeV. We also investigate the impact of the QCD corrections and the scale variations in searches of the anomalous $Wtb$ couplings. 
}
\end{abstract}

\maketitle

\section{Introduction}
\label{sec:intro}

The Large Hadron Electron Collider (LHeC)~\cite{AbelleiraFernandez:2012cc,Agostini:2020fmq,Mantysaari:2020aft} is a proposed facility of using a newly built electron beam of 60 GeV or higher energy to collide with the intense hadron beams of the LHC. As the high luminosity phase of the LHC will accumulate a total integrated luminosity of 3000$fb^{-1}$, the LHeC is expected to reach a total integrated luminosity of $100fb^{-1}$ . Such a programme will be devoted to probing the energy frontier and complementing the discovery potential of the LHC with measurements of deep inelastic scattering (DIS). It can be used to study the parton structure of the proton and  QCD dynamics~\cite{Rojo:2009ut,AbelleiraFernandez:2012ty,Cooper-Sarkar:2016udp,Paukkunen:2017phq,Caldwell:2018wqk,AbdulKhalek:2019mps,Abdolmaleki:2019tvj}, Higgs physics~\cite{Han:2009pe,Jager:2010zm,Biswal:2012mp,Esteves:2015nsa,Kumar:2015kca,Kumar:2015tua,Liu:2016ahc,Sun:2017mue,Mosomane:2017jcg,Flores-Sanchez:2018dsr,Hesari:2018ssq,DelleRose:2018ndz,Han:2018rkz,Li:2019jba,Cheung:2020ndx,Jueid:2021qfs}, trilinear couplings of gauge bosons~\cite{Forte:2015cia,Li:2017kfk}, top-quark physics~\cite{Cakir:2009xi,Cakir:2009rq,Dutta:2013mva,Cakir:2013cfx,Sarmiento-Alvarado:2014eha,TurkCakir:2017rvu,Sun:2018upk,Schwanenberger:2019bgo,Liu:2019wmi,Yang:2019uea,Rezaei:2020rwd} and new resonances~\cite{Tang:2015uha,Yang:2021skb}.

In particular the LHeC will provide a cleaner environment for the study of single-top production~\cite{Dutta:2013mva}. At the LHC, assuming a top-quark mass of 172.5 GeV, the top quark pair and single-top production cross sections are: $\sigma_{t\bar{t}}=984.5$ pb and $\sigma_{t+\bar{t}}=245$ pb at $\sqrt{s} = 14$ TeV~\cite{Czakon:2013goa,Berger:2017zof,Zyla:2020zbs}. The processes of single-top production contain $s$-, $t$- and $Wt$-channels  which are all related to $Wtb$ vertex. Theoretical efforts have been devoted to improving the theoretical predictions~\cite{Bordes:1994ki,hep-ph/9603265,hep-ph/9705398,hep-ph/9807340,hep-ph/0102126,hep-ph/0207055,Sullivan:2004ie,hep-ph/0408158,hep-ph/0510224,hep-ph/0512250,hep-ph/0504230,0907.4076,0903.0005,1007.0893,1012.5132,1010.4509,1103.2792,1102.5267,1207.5391,1210.7698,1305.7088,1404.7116,Assadsolimani:2014oga,1406.4403,1510.06361,Meyer:2016slj,Berger:2016oht,1603.01178,1708.09405,Carrazza:2018mix,1807.03835,1801.09656,Neumann:2019kvk,Kidonakis:2019nqa,1907.12586,Cao:2019uor,2005.12936,Campbell:2020fhf,Basat:2021xnn}. The single top quark production have been used to measure the CKM matrix element $Vtb$~\cite{Aaltonen:2015cra,Sirunyan:2020xoq} and to extract the top-quark mass~\cite{Khachatryan:2015hba,Alekhin:2016jjz,Aaboud:2018zbu}.  Besides, various new physics~\cite{He:1998ie,Tait:2000sh,Liu:2005dp,Zhang:2008yn,Gao:2009rf,Zhang:2010bm,Wang:2011uxa,Zhang:2011gh,Li:2011ek,Gao:2011fx,Wang:2012gp} have been searched for at the LHC with no discovery yet. For the LHeC, single top quark production via charged-current DIS is dominant in all the top quark production channels. We can utilize this unprecedented facility to measure the precise properties of the top quark and search for new physics.

In this work we present a fully differential calculation of the single top quark production at the LHeC in the 5-flavour scheme (5FS) at NLO in QCD. We focus on the leptonic decays of the top quark. The hadronic decays will be affected more by various non-perturbative effects of QCD and also the SM backgrounds. We include full off-shell and non-resonant contributions in our calculation.  We study the QCD corrections to various differential distributions in a typical fiducial  region of the LHeC. Especially we show their impact to the precision measurement of the top quark mass and the searches of the anomalous $Wtb$ couplings.

The rest of our paper is organized as follows. In Sec.$~\ref{sec:Calculation}$, we present our theoretical framework and the numerical results of total cross sections. In Sec.$~\ref{sec:Analysis}$, we show the numerical results for differential distribution, and the implications to the measurement of the top quark mass and the searches of the anomalous $Wtb$ couplings. Finally, in Sec.$~\ref{sec:Conclusions}$, we present our conclusion.

\section{NLO Calculation}
\label{sec:Calculation}
\begin{figure}[t!]
	\centering
	\includegraphics[scale=0.25]{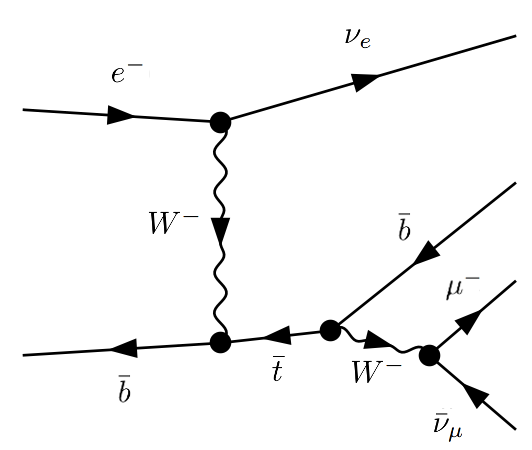}
	\includegraphics[scale=0.25]{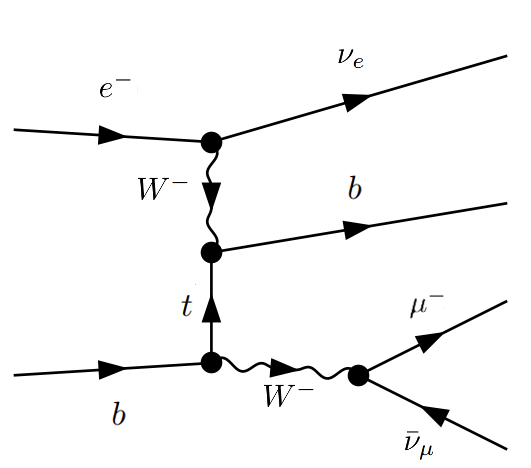}
	\caption{\label{fig:1} The LO Feynman diagrams for the single top quark production with leptonic decays at the LHeC. We also include the subprocess that has indistinguishable final-state as the single top quark production.}
\end{figure}

In this section, we will describe the major theoretical details and numerical results for the total cross sections. The generic processes under consideration are $e^{-} \bar{b} \rightarrow \nu_e \bar{t} \rightarrow \nu_e \ell^- \bar{\nu}_{\ell^-} \bar{b}$ and $e^{-} {b} \rightarrow  \nu_e \ell^- \bar{\nu}_{\ell^-} {b}$. The LO Feynman diagrams are depicted in Fig.~\ref{fig:1}. Our computation is based on combination of \texttt{GoSam2.0}~\cite{Cullen:2011ac, Cullen:2014yla} and a native Monte Carlo framework. We use \texttt{GoSam2.0} program to generate one-loop results for the virtual amplitudes. We adopt the dipole subtraction method~\cite{Catani:1996vz,Catani:2002hc} to construct local subtraction terms and their integrations for real emissions. All these ingredients are embedded into a MC program which can accomplish the cancellation of IR singularities and phase-space integrations.

In the following we introduce the dipole subtraction method used to deal with NLO calculations of subprocess in single top quark production. We present the numerical result for the total cross section at the NLO based on the framework. The discussions on scale variations at both LO and NLO are shown at the end of this section.

\subsection{Theoretical framework}

We use the 5FS which treats the bottom quark as a massless parton in initial hadrons.  The 5FS ensures a resummation of large quasi-collinear logarithms from gluon splitting to bottom quarks through the parton distribution function of bottom quarks. The NLO cross section can be written as\cite{Catani:1996vz,Catani:2002hc}
\begin{equation}
	\label{eq:1}
	\begin{aligned}
		&\sigma^{\mathrm{NLO}\{5\}}(p)+\sigma^{\mathrm{NLO}\{4\}}(p)+\int_{0}^{1} d x \hat{\sigma}^{\mathrm{NLO}\{4\}}(x ; x p)\\
		\equiv&\int_{5}\left[\left(\mathrm{~d} \sigma^{\mathrm{R}}(p)\right)_{\epsilon=0}-\left(\sum_{\text {dipoles }} \mathrm{d} \sigma^{\mathrm{B}}(p) \otimes \mathrm{d} V_{\text {dipole }}\right)_{\epsilon=0}\right] \\
		&+\int_{4}\left[\mathrm{~d} \sigma^{\mathrm{V}}(p)+\mathrm{d} \sigma^{\mathrm{B}}(p) \otimes \boldsymbol{I}\right]_{\epsilon=0}\\
		&+\int_{0}^{1} d x \int_{4}\left[\mathrm{~d} \sigma^{\mathrm{B}}(x p) \otimes(\boldsymbol{P}+\boldsymbol{K})(x)\right]_{\epsilon=0},
	\end{aligned}
\end{equation}
where the contributions $\sigma^{\mathrm{NLO}\{5\}}(p)$ and $\sigma^{\mathrm{NLO}\{4\}}(p)$ (with 5-body final state and 4-body final state, respectively) represent the subtracted real contributions and virtual contributions including integrated dipoles. The third term $\int_{0}^{1} d x \hat{\sigma}^{\mathrm{NLO}\{4\}}(x ; x p) $is a finite remainder which comes from the cancellation of the $\epsilon$ poles of the collinear counter-terms. It contains an additional one-dimensional integration
with respect to the longitudinal momentum fraction $x$. The $\boldsymbol{P}$ and $\boldsymbol{K}$ are universal functions of $x$ which are finite for $\epsilon \rightarrow 0$. The $\mathrm{d}\sigma^{\mathrm{R}}$, $\mathrm{d} \sigma^{\mathrm{V}}$ and $\mathrm{d} \sigma^{\mathrm{B}}$ are the fully differential cross section from real, virtual(one-loop) and Born contributions. The dipole factors $\mathrm{d} V_{\text {dipole }}$ describe the two-parton decays of the emitters. The factor $\boldsymbol{I}$ is derived from the dipole factors by integrating out a single parton phase space, which will cancel the $\epsilon$ poles in virtual contributions.

\begin{figure}[t!]
	 \centering
	\includegraphics[scale=0.25]{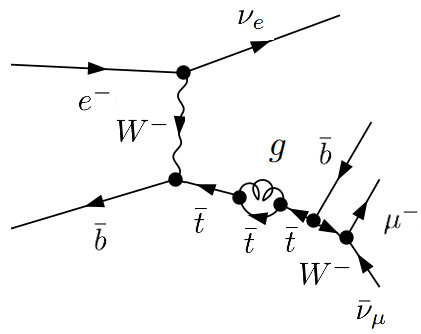}
	\includegraphics[scale=0.25]{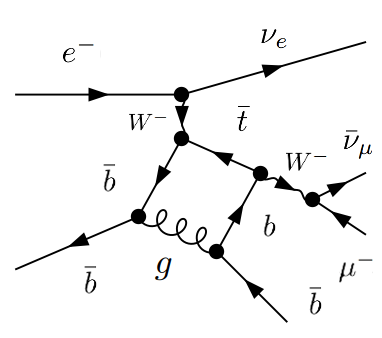}
	\includegraphics[scale=0.25]{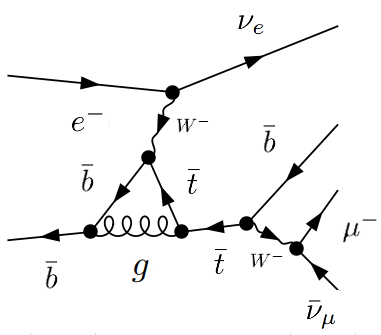}
	\includegraphics[scale=0.25]{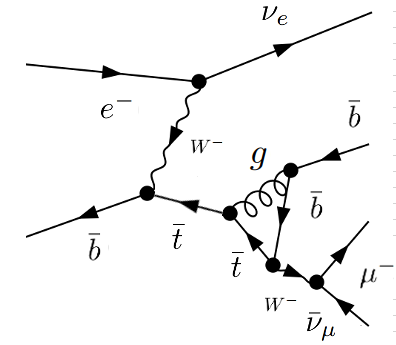}
	\caption{\label{fig:2} One-loop Feynman diagrams for the single top quark production with leptonic decays at the LHeC with a $\bar{b}$ quark in the initial-state.}
\end{figure}

\begin{figure}[t!]
	\centering
	\includegraphics[scale=0.2]{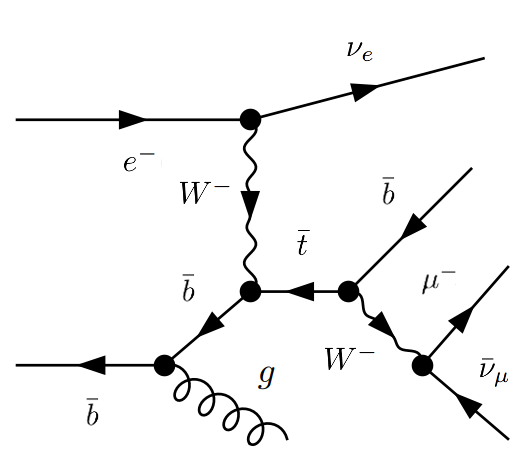}
	\includegraphics[scale=0.2]{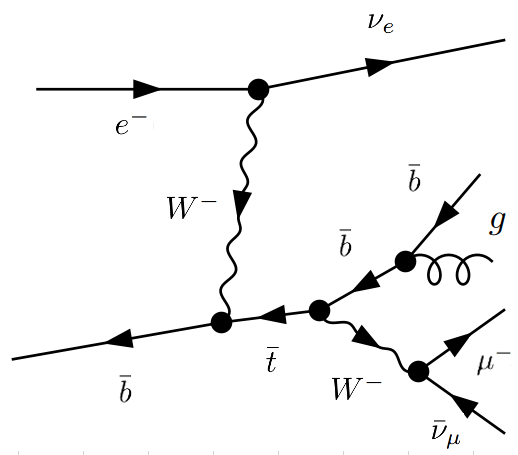}
	\includegraphics[scale=0.2]{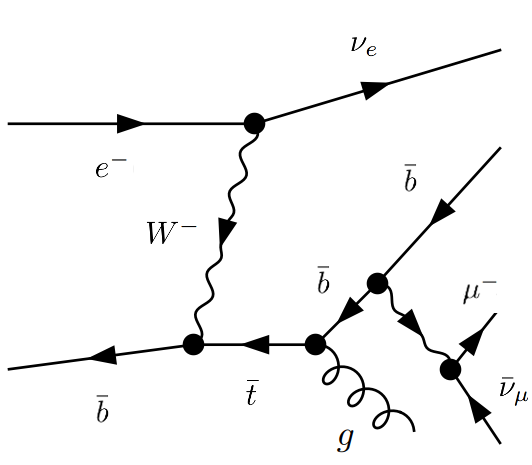}
	\caption{\label{fig:3} Real emission Feynman diagrams for the single top quark production with leptonic decays at the LHeC with a $\bar{b}$ quark in the initial-state.}
\end{figure}

\begin{figure}[t!]
	\centering
	\includegraphics[scale=0.18]{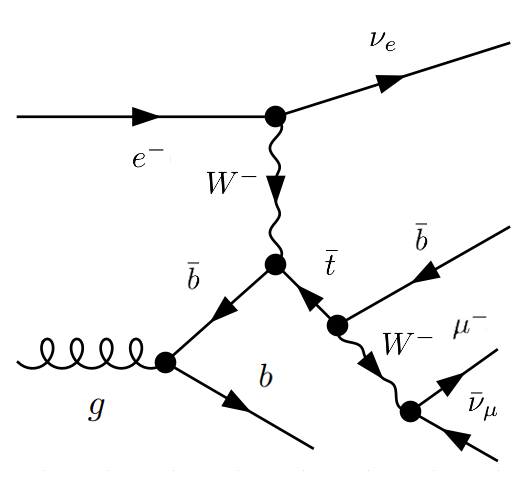}
	\hspace{0.3in}
	\includegraphics[scale=0.18]{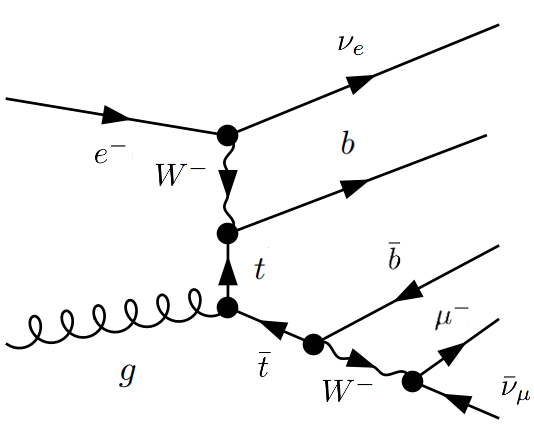}
	\hspace{0in}
	\includegraphics[scale=0.18]{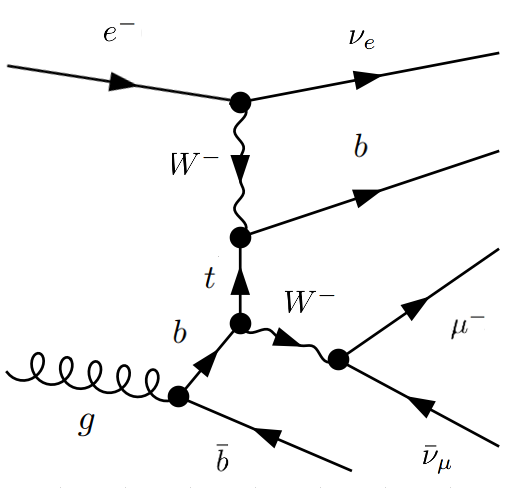}
	\caption{\label{fig:4} Real emission Feynman diagrams for the single top quark production with leptonic decays at the LHeC with a gluon in the initial-state.}
\end{figure}

We show the one-loop Feynman diagrams with a $\bar{b}$ quark in the initial state in Fig.~\ref{fig:2}. There are similar one-loop diagrams for the process with a $b$ quark in the initial-state which are not shown for simplicity. The real emission Feynman diagrams with a $\bar{b}$ quark or gluon in the initial state are shown in Fig.~\ref{fig:3} and~\ref{fig:4} respectively. The corresponding subtraction terms include those of initial-state emitter and final-state spectator $\mathcal{D}_{j}^{a i}$ and$/$or final-state emitter and initial-state spectator $\mathcal{D}_{i j}^{a}$
\begin{widetext}
	\begin{equation}
		\begin{array}{l}
			\mathcal{D}_{j}^{a i}\left(p_{1}, \ldots, p_{5} ; p_{a}, \ldots\right)= \\
			\quad-\frac{1}{2 p_{a} p_{i}} \frac{1}{x_{i j, a}}_{4, a i}	\left\langle\ldots, \tilde{j}, \ldots ; \widetilde{a i}, \ldots\left|\frac{\boldsymbol{T}_{j} \cdot \boldsymbol{T}_{a i}}{\boldsymbol{T}_{a i}^{2}} \mathbf{V}_{j}^{a i}\right| \ldots, \tilde{j}, \ldots ; \widetilde{a i}, \ldots\right\rangle_{4, \widetilde{a i}},
		\end{array}
	\end{equation}
	
	\begin{equation}
		\begin{array}{l}
			\mathcal{D}_{i j}^{a}\left(p_{1}, \ldots, p_{5} ; p_{a}, \ldots\right)= \\
			\quad-\frac{1}{2 p_{i} p_{j}} \frac{1}{x_{i j, a}} _{4, a}\left\langle\ldots, \tilde{i j}, \ldots ; \tilde{a}, \ldots\left|\frac{\boldsymbol{T}_{a} \cdot \boldsymbol{T}_{i j}}{\boldsymbol{T}_{i j}^{2}} \mathbf{V}_{i j}^{a}\right| \ldots, \tilde{i j}, \ldots ; \tilde{a}, \ldots\right\rangle_{4, a},
		\end{array}
	\end{equation}
	
	\begin{equation}
		x_{i j, a}=\frac{p_a p_i+p_a p_j-p_i p_j}{p_a p_i+p_a p_j}.
	\end{equation}
\end{widetext}

The $\mathcal{D}_{j}^{a i}$ terms cancel the singularities of the matrix elements when the final-state parton $i$ and the initial-state parton a become collinear. Similarly, the $\mathcal{D}_{i j}^{a}$ cancel the singularity of the matrix elements when the final-state gluon and $b$ or $\bar{b}$ become collinear. $\boldsymbol{T}$ and $\boldsymbol{V}$ are colour charge operator and spin function respectively. The $| \ldots \rangle_{4, \widetilde{a i}}$ and $| \ldots \rangle_{4, a}$ are reduced born matrix elements by replacing the parton pair $a,i$ with a single parton $\tilde{ai}$ and by replacing the parton pair $i,j$ with a single parton $\tilde{i j}$ respectively.

The calculations are carried out in the complex-mass scheme~\cite{Aeppli:1993rs,Denner:1999gp} such that our results are valid in all regions of the phase space including when the top quark is off-shell. The complex-mass scheme is a generalization of the on-shell renormalization scheme, in which case the bare top quark mass includes a complex renormalized mass $\mu_{t}^{2}=m_{t}^{2}-i m_{t} \Gamma_{t}$ and a complex counter-term $\delta \mu_{t}$. The cross sections are dominated by contributions from diagrams with a top quark resonance. However, we have also included those non-resonant diagrams as can be seen from Figs.~\ref{fig:1}-\ref{fig:4}.

We are now ready to calculate predictions for any infrared and collinear safe observables provided with the virtual corrections from  \texttt{GoSam2.0} and the formulas of the dipole terms in~\cite{Catani:1996vz,Cullen:2011ac}. As for the parameterization of the  phase space, we use the multi-channel approach to accommodate for the singular structures of the top-quark resonance. The numerical integrations are performed with the Monte Carlo library \texttt{Cuba}~\cite{Hahn:2016ktb}.

\subsection{Numerical Result}
\label{sec:num}
We study the single top quark production in $ep$ collisions at the LHeC with an electron beam energy of 70 GeV and a proton beam energy of 7 TeV. We use the following set of the SM parameters in the numerical calculations~\cite{Zyla:2020zbs}

\begin{equation}
	\begin{array}{l}
		m_{Z}=91.1876 \,\mathrm{GeV}, m_{t}=172.5 \,\mathrm{GeV},\\ 
		m_{W}=80.385 \,\mathrm{GeV},G_{\mathrm{F}}=1.16639 \times 10^{-5}\,\mathrm{GeV^{-2}}.
	\end{array}
\end{equation}
The  CT18 NNLO parton distribution functions (PDFs)~\cite{Hou:2019efy}  and the strong coupling constant $\alpha_s(m_Z)=0.118$ are used throughout all the calculations. The nominal choice of the factorization and renormalization scales are $\mu_{R}=\mu_{F}=m_{t}/2 $, and the scale variations are calculated by varying the two scales simultaneously from $m_{t}/4$ to $m_{t}$.

We list our predictions of the inclusive cross sections for the single top quark production with leptonic decays at different perturbative orders in Table~\ref{tab:1}, with scale variations shown in percentages. We find that the LO cross section is dominated by the subprocess with a bottom anti-quark in the initial state. The NLO QCD corrections reduce the total cross sections by 8.5\%. The full NLO corrections consist of three pieces from subprocesses with the bottom anti-quark, bottom quark and gluon in the initial state. The NLO corrections are also dominated by the subprocess with a bottom anti-quark in the initial state. The scale variations at NLO are decreased by 5 times comparing with LO, and are within 3\%.

\begin{table}[t]
	\centering
	\begin{tabular}{|c|c|c|}
		\hline
		inclusive [pb]&LO&NLO\\
		\hline 
		\hline 
		$\sigma[total]$  & $0.281^{+8.2\%}_{-11\%}$&$0.257^{+0.92\%}_{+2.6\%}$\\
		
		$\sigma[\,\bar{b} \,]$  & $0.281$&0.264\\
		
		$\sigma[\,b\,]$  & $5.35\times10^{-4}$&$5.18\times10^{-4}$\\
		
		$\sigma[\,g\,]$  & &$-6.97\times10^{-3}$\\
		\hline
	\end{tabular}

	\caption{\label{tab:1}  Inclusive cross sections for the single top quark production with leptonic decays at the LHeC at various orders in QCD with a nominal scale choice of $m_t/2$. The scale variations are calculated by varying the scales from $\mu_F = \mu_R = m_t/4$ to $m_t$, and are shown in percentages. In the numbers of cross sections the upper(lower) variation corresponds to the scale choice of $m_t$($m_t/4$). Separate contributions from three subprocesses with different initial state are also shown.}
\end{table}

For the single top quark production, we apply various selection cuts to account for the finite kinematic coverage of the detectors and to suppress the SM background from the associated production of vector bosons with jets. Final-state quarks and gluons are clustered into jets using the anti-k$_{\rm{T}}$ jet algorithm~\cite{Cacciari:2008gp}. The jet-resolution parameter is set to 0.4. We require at least one $b$-tagged jet in the final state. The cuts on the transverse momentum and pseudorapidity of the jet and lepton are shown below

\begin{equation}
	\begin{array}{ll}
		\left|\eta_{\ell^-}\right|<5, & p_{T, \ell^-}>10 \,\mathrm{GeV}\\
		\left|\eta_{j}\right|<5, &p_{T, j e t}>30 \,\mathrm{GeV}.\\
	\end{array}\label{cuts}
\end{equation}

We show the predictions of the fiducial cross sections at LO and NLO in Table~\ref{tab:2}. We list contributions from the different subprocesses and show the scale variations in percentages.
We find that the selection cuts reduce the cross section by about 14\% at LO by comparing to Table~\ref{tab:1}. The QCD corrections are still dominated by contributions from the subprocess with a $\bar{b}$ quark in the initial state. The full NLO corrections reduce the fiducial cross section by 15\%. The scale variations of the NLO predictions are about 3\%, largely reduced comparing to the LO ones. We can also calculate the efficiency which is defined as the ratio of the fiducial cross section to the inclusive cross section. From the results in Tables~\ref{tab:1} and~\ref{tab:2}, we derive the efficiency as 0.861 and 0.798 at LO and NLO respectively.

\begin{table}[t]
	\begin{tabular}{|c|c|c|}
		\hline
		fiducial [pb]&LO&NLO\\
		\hline 
		\hline
		$\sigma[total]$  & $0.242^{+8.1\%}_{-11\%}$&$0.205^{+0.76\%}_{+3.1\%}$\\
		
		$\sigma[\,\bar{b} \,]$  &0.242 &0.207\\
		
		$\sigma[\,b\,]$  &5.01$\times10^{-4}$&$4.62\times10^{-4}$\\
		
		$\sigma[\,g\,]$  & &$-2.95\times10^{-3}$\\
		\hline
	\end{tabular}
	\caption{\label{tab:2}  Similar to Table~\ref{tab:1} for fiducial cross sections.}
\end{table}

\begin{figure}[t!]
	\centering
	\includegraphics[scale=0.4]{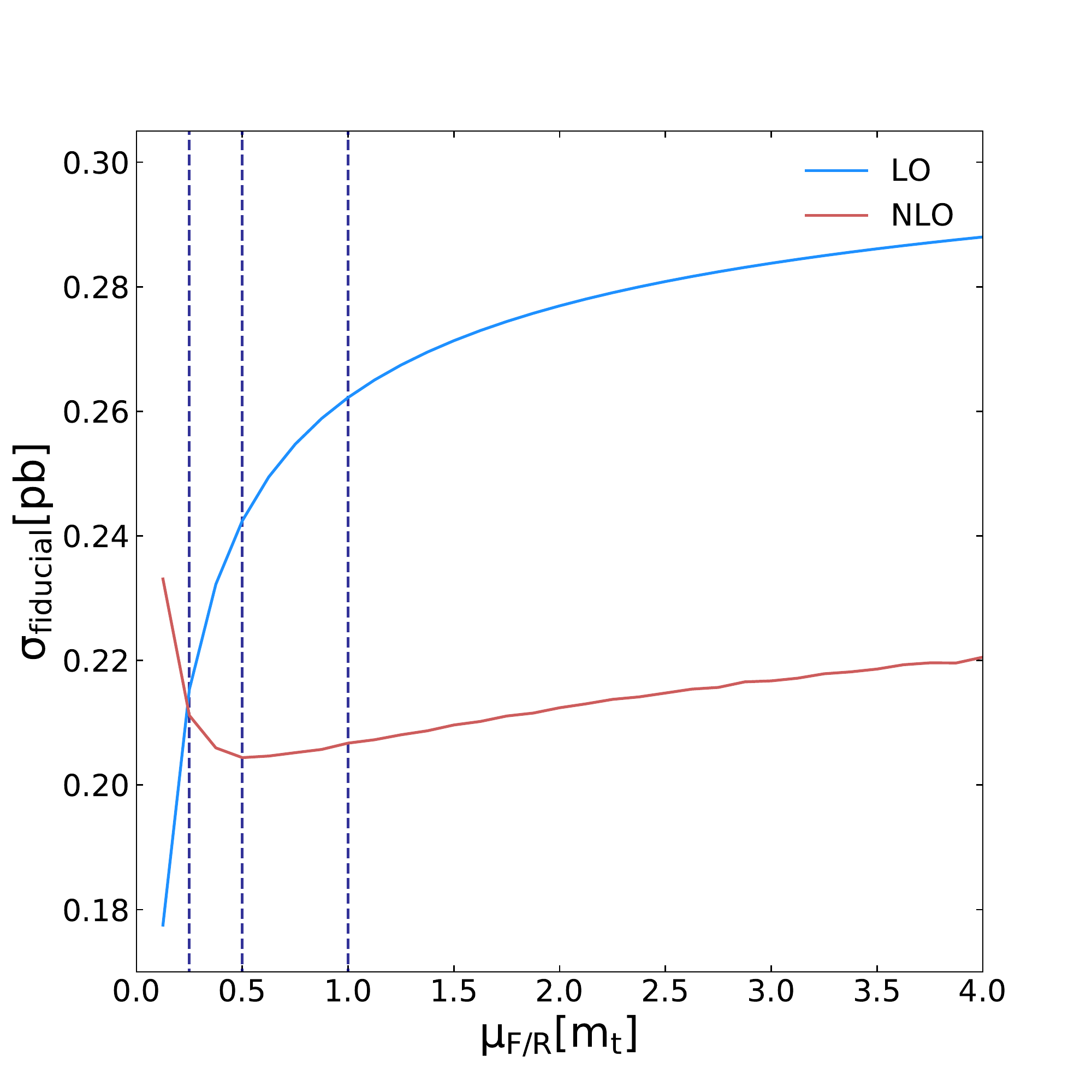}
	\caption{\label{fig:fidu} The dependence of the fiducial cross section on the factorization and renormalization scales ($\mu_{F}=\mu_{R}$)}
\end{figure}

The dependence of the fiducial cross section on the factorization and renormalization scales ($\mu_{F}=\mu_{R}$) are shown in Figure~\ref{fig:fidu}. The three vertical lines are corresponding to scales $m_t/4$, $m_t/2$ and $m_t$ respectively. The two curves denote cross sections at the LO and NLO. The cross sections change dramatically below the scale $m_t/4$. We find that the cross sections are more stable against the scale choice in the range $m_t/4 < \mu_{F/R} < m_t$. That motivates our nominal choice of the scale and its variation range.

Table~\ref{tab:scut} shows the predictions of the fiducial cross sections at the LO and NLO with more stringent cuts shown below
\begin{equation}
	\begin{array}{ll}
		\left|\eta_{\ell^-}\right|<3.5, & p_{T, \ell^-}>25 \,\mathrm{GeV}\\
		\left|\eta_{j}\right|<3.5, &p_{T, j e t}>50 \,\mathrm{GeV}.\\
	\end{array}
\end{equation}
The cuts reduce the cross section by about 53\% at the LO by comparing to Table~\ref{tab:1}. The scale variations of  the LO prediction are identical to those shown in Table~\ref{tab:1}. The NLO corrections reduce the fiducial cross section by about 18\% with scale variations slightly larger than those in Table~\ref{tab:2}. We use the selection cuts in Eq.~(\ref{cuts}) as our default setup in the rest of the paper.

\begin{table}[t]
	\centering
	\begin{tabular}{|c|c|c|}
		\hline
		fiducial [pb]&LO&NLO\\
		\hline 
		\hline
		$\sigma[total]$  & $0.132^{+8.2\%}_{-11\%}$&$0.108^{+0.14\%}_{+4.2\%}$\\
		
		$\sigma[\,\bar{b} \,]$  &0.132 &0.104\\
		
		$\sigma[\,b\,]$  &4.08$\times10^{-4}$&$3.63\times10^{-4}$\\
		
		$\sigma[\,g\,]$  & &$3.71\times10^{-3}$\\
		\hline
	\end{tabular}
	\caption{\label{tab:scut}  Similar to Table~\ref{tab:1}  for fiducial cross sections with more stringent cuts.}
\end{table}

\section{Phenomenological Analysis}
\label{sec:Analysis}
In this section, we present a phenomenological analysis of the differential distributions based on our MC program. We show the QCD corrections and theoretical uncertainties of the distributions. We propose a method of determining the top-quark mass using the average transverse momentum of the charged lepton, and estimate various uncertainties of the measurement. We also show impact of the QCD corrections on constraining the anomalous couplings of the top quark.

\subsection{LO and  NLO Predictions}

\begin{figure}[t!]
	\centering
	\includegraphics[scale=0.47]{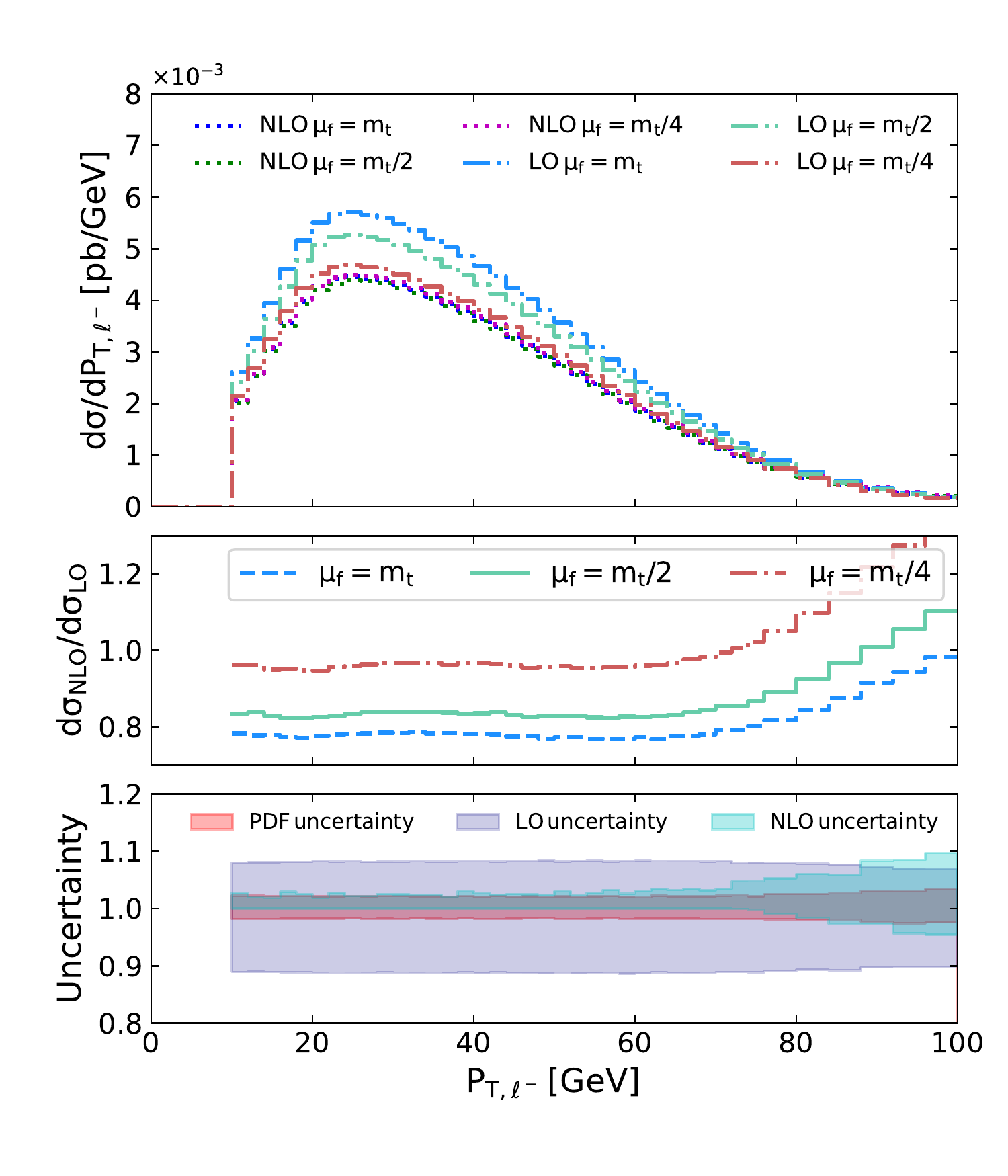}
	\includegraphics[scale=0.47]{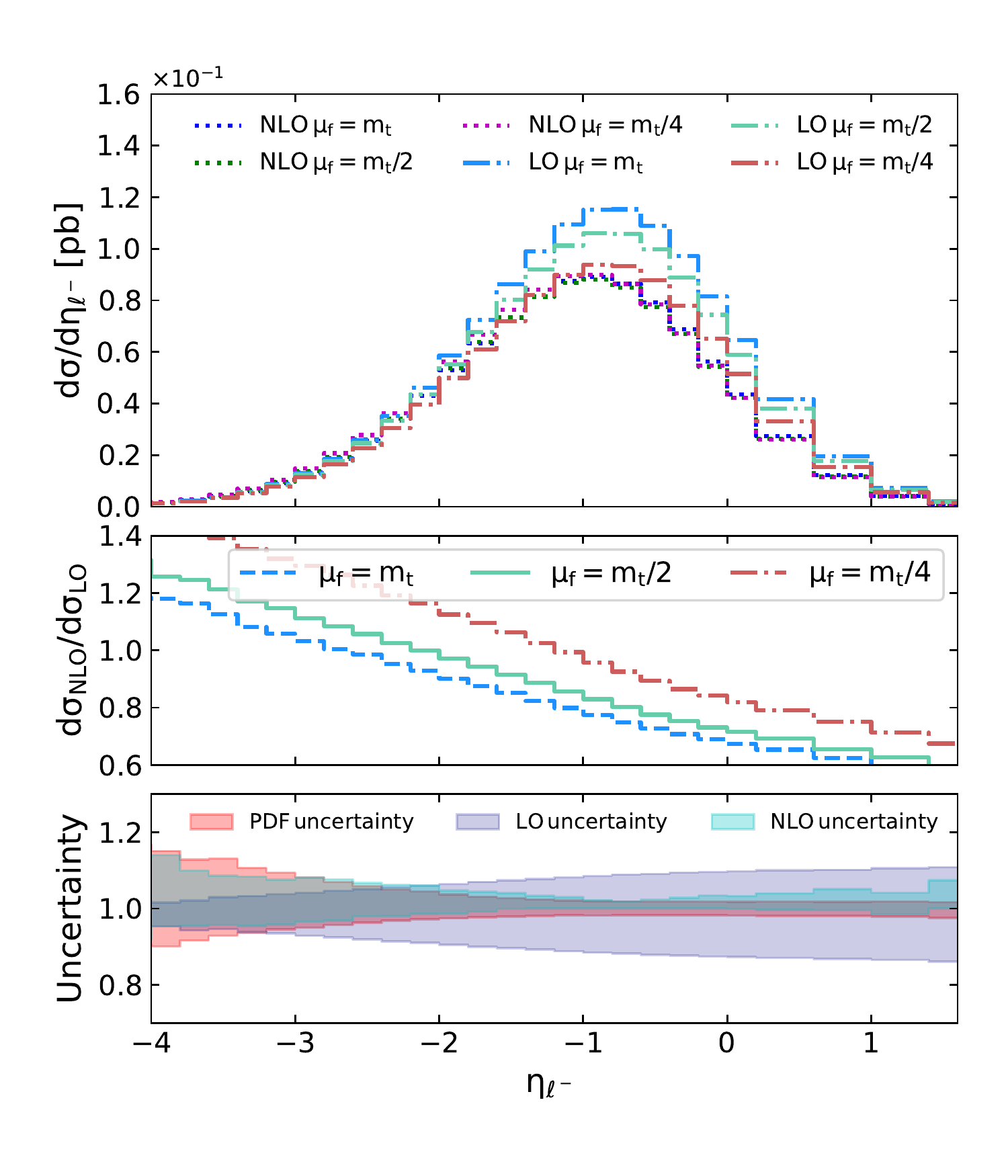}
	\caption{\label{fig:5} Transverse momentum and pseudorapidity distributions of the charged lepton with fiducial cuts applied at LO and NLO with a nominal scale choice of $m_t/2$ and alternative scale choices of $m_t/4$ and $m_t$. The middle panel shows the ratio of the NLO predictions to LO predictions ($d\sigma_{NLO}/d\sigma_{LO}$) with each scale choice. The lower plot shows the PDF uncertainties and scale variations at both LO and NLO.}
\end{figure}

\begin{figure}[t!]
	\centering
	\includegraphics[scale=0.47]{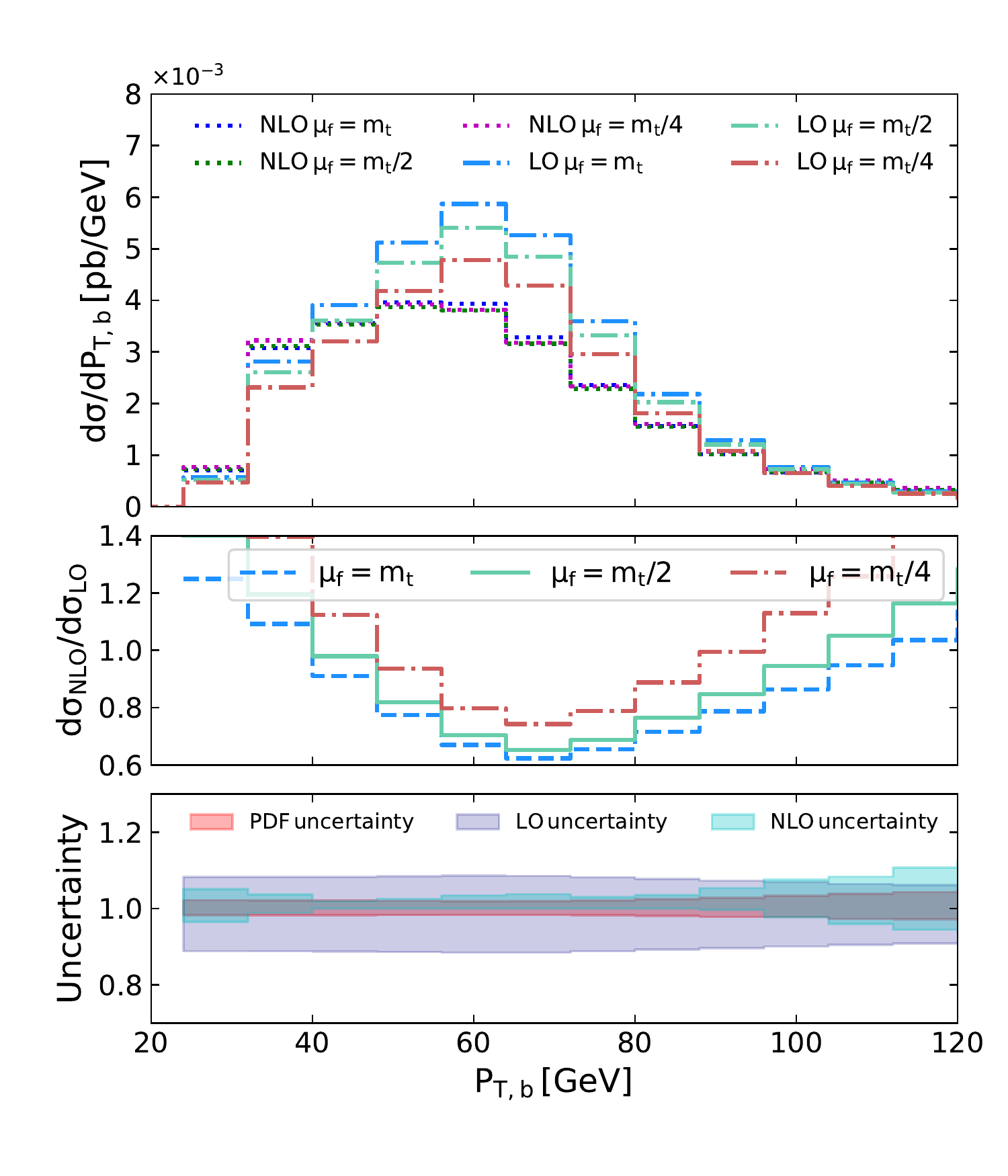}
	\includegraphics[scale=0.47]{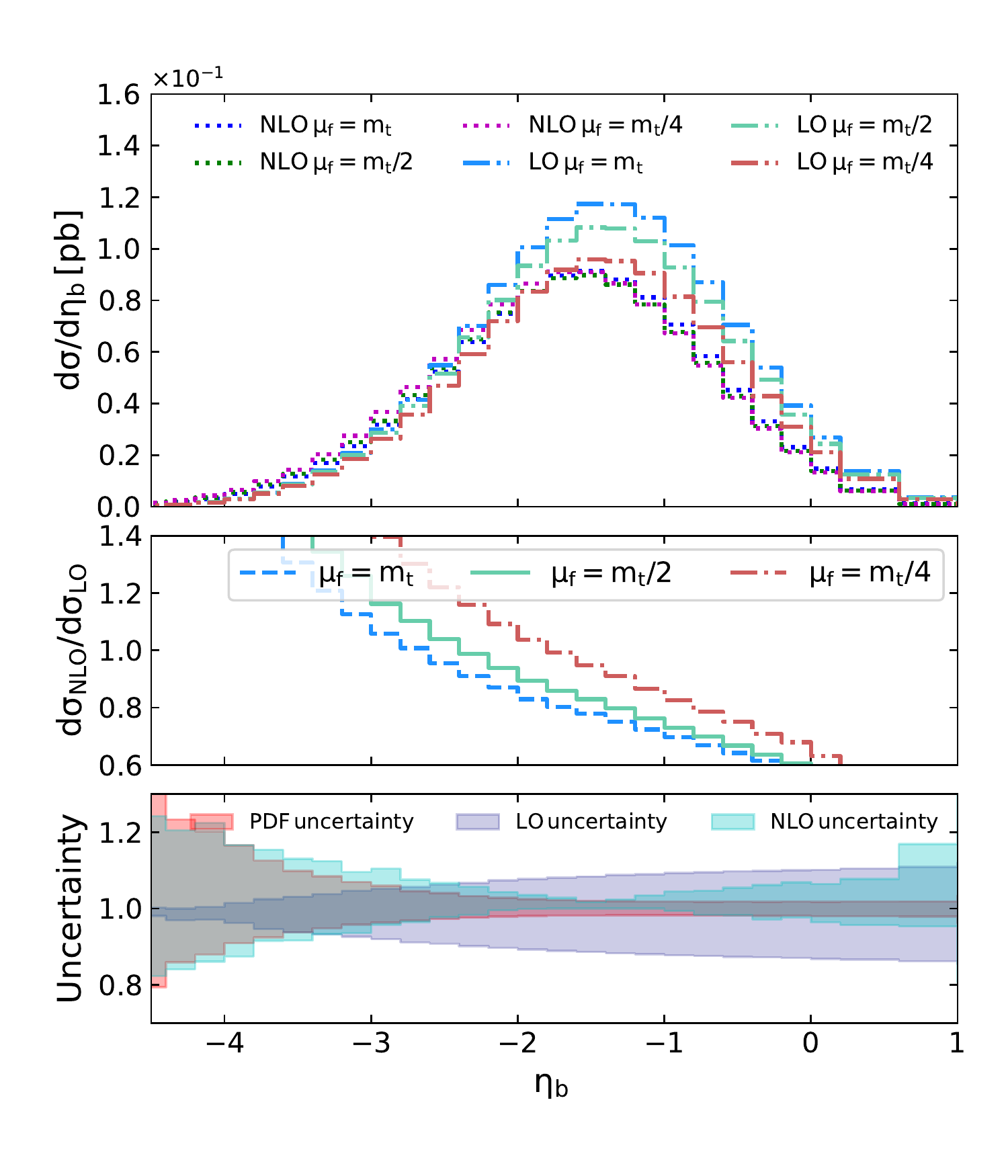}
	\caption{\label{fig:6} Similar to Figure~\ref{fig:5} for the transverse momentum and pseudorapidity distributions of the b-jet.}
\end{figure}

\begin{figure}[t!]
	\centering
	\includegraphics[scale=0.47]{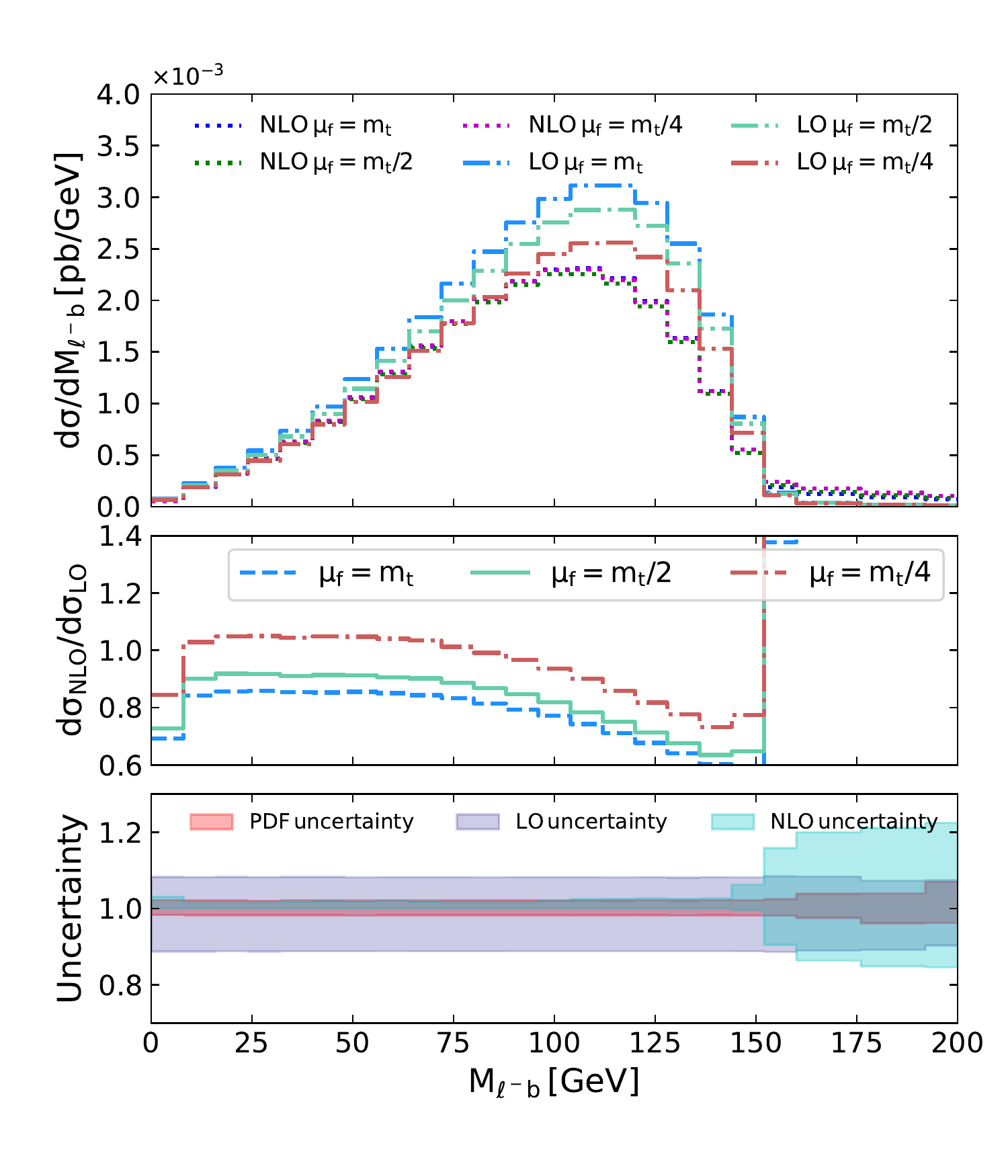}
	\caption{\label{fig:7} Similar to Figure~\ref{fig:5} for the invariant mass distributions of the charged lepton and b-jet system ($M_{\ell^-b}$).}
\end{figure}

\begin{figure}[t!]
	\centering
	\includegraphics[scale=0.47]{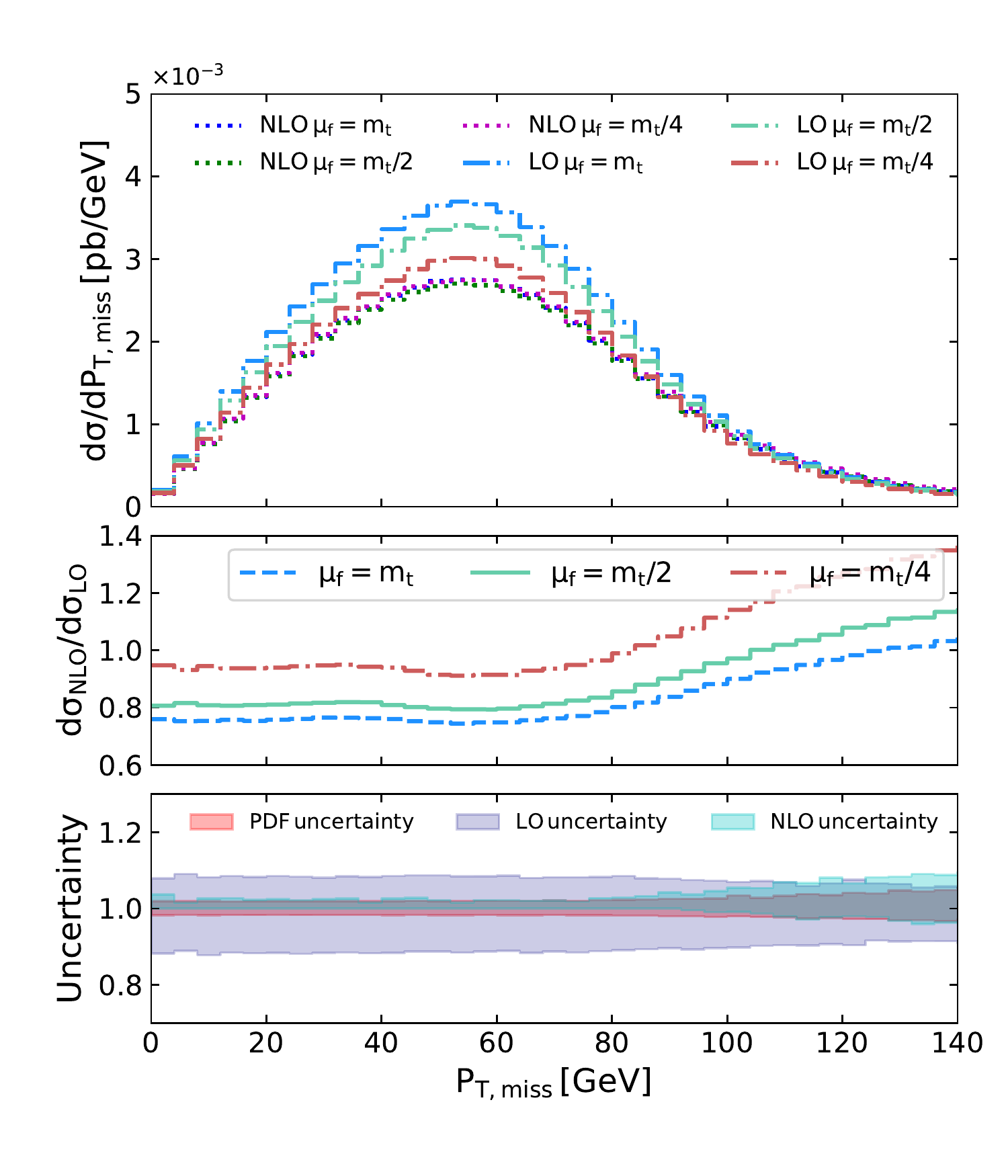}
	\caption{\label{fig:8} Similar to Figure~\ref{fig:5} for the distribution of the missing transverse momentum ($P_{T,miss}=|P_{T,\nu_{\ell^-}}+P_{T,\bar{\nu}_e}|$) distribution. }
\end{figure}

\begin{figure}[t!]
	\centering
	\includegraphics[scale=0.47]{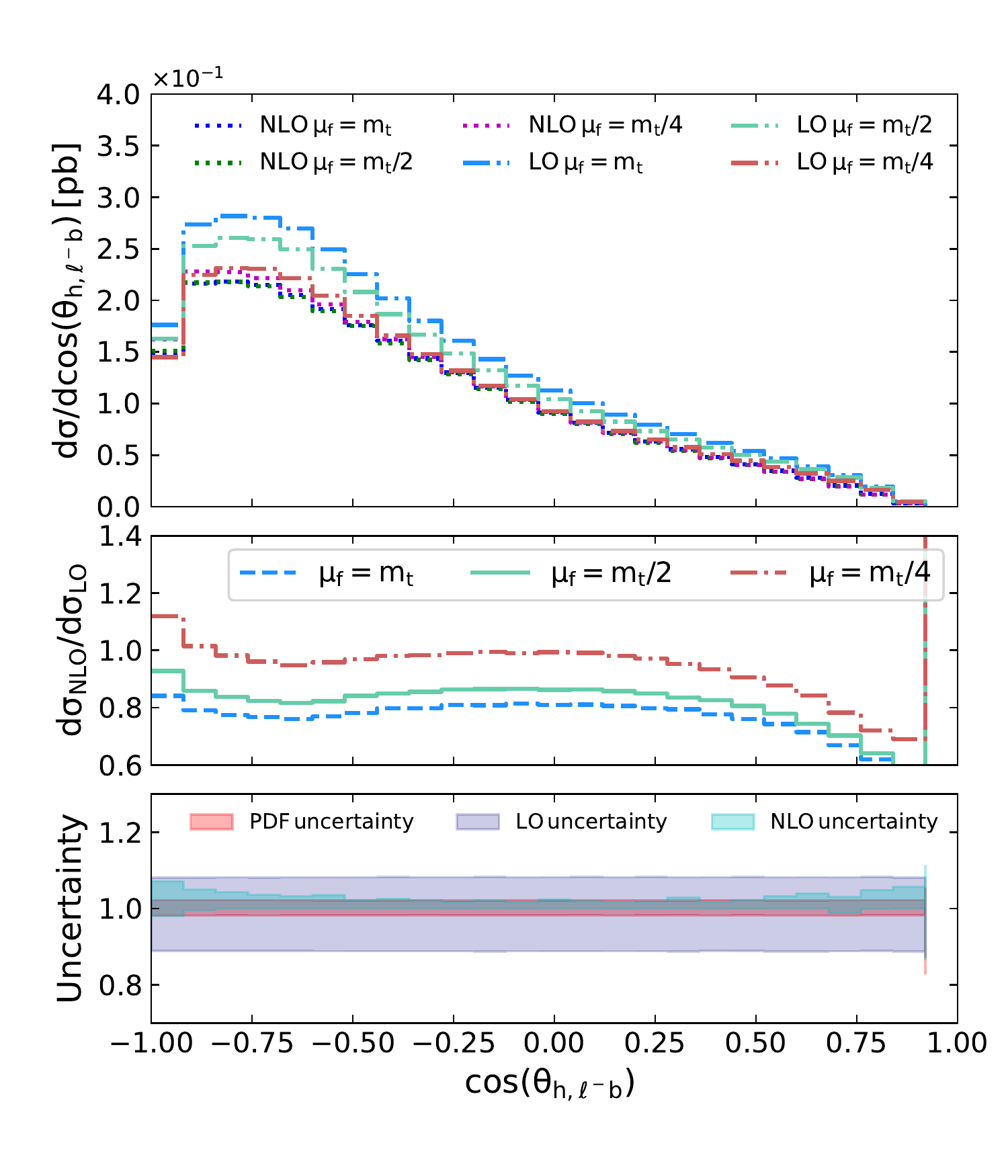}
	\includegraphics[scale=0.47]{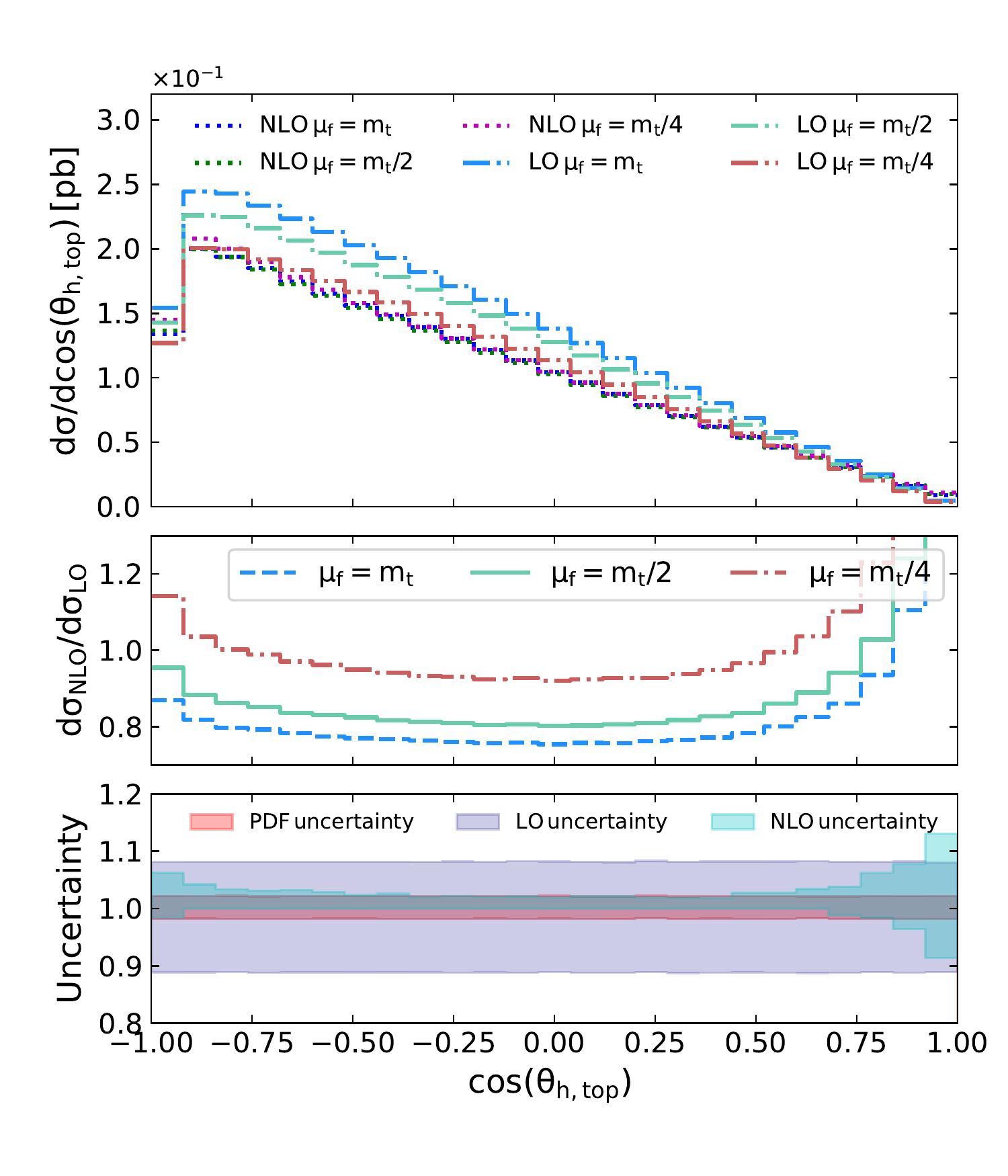}
	\caption{\label{fig:9} Similar to Figure~\ref{fig:5} for the distributions of helicity angles of the visible-particles-reconstructed top-quark and the truth top-quark.}
\end{figure}

We present predictions of various distributions within the fiducial region defined in Sec.$~\ref{sec:num}$ in Figs.$~\ref{fig:5}-\ref{fig:9}$. For the plot of each distribution, the upper panel shows the LO and NLO distributions with a nominal scale choice of $m_t/2$ and alternative scale choices of $m_t/4$ and $m_t$. The middle panel shows the ratio of the NLO predictions to LO predictions ($d\sigma_{NLO}/d\sigma_{LO}$) with each scale choice. The lower panel shows the PDF uncertainties and scale variations at both LO and NLO. The  PDF uncertainties at 68\% CL are calculated at LO using the 58 error PDF sets in the CT18 NNLO PDFs~\cite{Hou:2019efy}, and are normalized to the LO distributions with the scale choice $m_t/2$. The scale variations of the LO(NLO) predictions are also shown in the lower panel, and are normalized to the LO(NLO) predictions with the scale choice $m_t/2$.

The distribution of the transverse momentum of the charged lepton is presented in Figure$~\ref{fig:5}$. In the upper panel,  the peak of the distribution occurs at around 25 GeV, and the $P_T$  of the charged lepton can extend to 100 GeV. The NLO corrections decrease the normalization of the distribution without changing the position of the peak. The NLO distributions of three different scales are closer to each other than the LO ones
because of the weaker scale dependence  of the NLO predictions. From the middle panel we can see a better perturbative convergence for the scale of $m_t/4$ in the low $P_T$ region, where the ratio $d\sigma_{NLO}/d\sigma_{LO}$ is close to one. In the lower panel, the scale variations at LO are much larger than the PDF uncertainties. The NLO corrections reduce the scale variations significantly to a level that is comparable to or even smaller than the PDF uncertainties. Figure$~\ref{fig:5}$ also shows the pseudorapidity distribution of the charged lepton. The peak of the distribution occurs at a negative value due to the asymmetric collision. For all three scale choices the ratio of NLO to LO distribution decreases with the pseudorapidity. The scale variations at NLO are close to the PDF uncertainties in the full range of pseudorapidity.

We show the transverse momentum and pseudorapidity distributions of the b-jet in Figure~\ref{fig:6}. In the transverse momentum distribution, the peak of the distribution occurs at around 60 GeV. The NLO corrections decrease the normalization of the transverse momentum distribution and shift the peak position to lower $P_T$. The ratios of the NLO to the LO distribution show a minimum at $P_T$ close to the peak region. These behaviors are due to the non-resonant contributions and the hard gluon radiations. Pseudorapidity  distributions of the b-jet are similar to the distributions of the charged lepton. They peak at negative values of $\eta_{b}$. However, the ratios of the NLO to the LO predictions in the middle panel decrease more dramatically than in the cases of the charged lepton. The scale variations are largely reduced at the NLO in low $|\eta_{b}|$ region. For large  $|\eta_{b}|$ the scale variations at the LO are severely underestimated.

Figure$~\ref{fig:7}$ depicts distributions of the invariant mass  from the system of the charged lepton and the b-jet (the visible decay products of the top quark). The sharp cutoff in the distribution is due to the kinematic constraint $M_{\ell^-b}^2<M_t^2-M_W^2\approx(150\,$ GeV$)^2$. The ratio of the NLO to the LO distribution increase right after the cutoff because of the non-resonant and off-shell effects at the NLO.  This distribution has been used to measure the top-quark mass in the single top quark production at the LHC~\cite{Esch:2014bqa,CMS:2017mpr}. We present the distribution of the total missing transverse momentum ($P_{T,miss}=|P_{T,\nu_{\ell^-}}+P_{T,\bar{\nu}_e}|$) in Figure~\ref{fig:8}. The peak of the distribution occurs at a much larger $P_T$ compared with the distribution of $P_{T,\ell^-}$. The ratio $d\sigma_{NLO}/d\sigma_{LO}$ rises significantly in the tail region. The PDF uncertainties and scale variations at the NLO are much smaller than scale variations at the LO. 

For the leptonic decays of the top quark, the charged lepton $\ell^-$ is strongly correlated with the spin direction of  the top quark. We define the helicity angle $\theta_h$ as the angle between the direction of the momentum of the charged lepton and the spin of the top quark in the rest frame of the top quark~\cite{Czarnecki:1994pu,Aubert:2006dc}. In the single top quark production, the spin of the top anti-quark is always in the opposite direction of the momentum of the incident electron. We show the distributions of the helicity angles in Figure~\ref{fig:9}. The $cos(\theta _{h,\ell^-b})$ is the cosine of the helicity angle based on the top quark reconstructed with only visible decay products and the $cos(\theta _{h,top})$ based on the truth top quark. The $cos(\theta _{h,top})$ distribution displays a strong correlation pattern except in the extreme backward region where the fiducial cuts play an important role. The correlations are weakened in the distribution of $cos(\theta _{h,\ell^-b})$ due to the approximation with the reconstructed top quark. The ratios  $d\sigma_{NLO}/d\sigma_{LO}$ are flat in most regions of the helicity angles.

\subsection{Top Quark Mass}

\begin{figure}[t!]
	\includegraphics[scale=0.45]{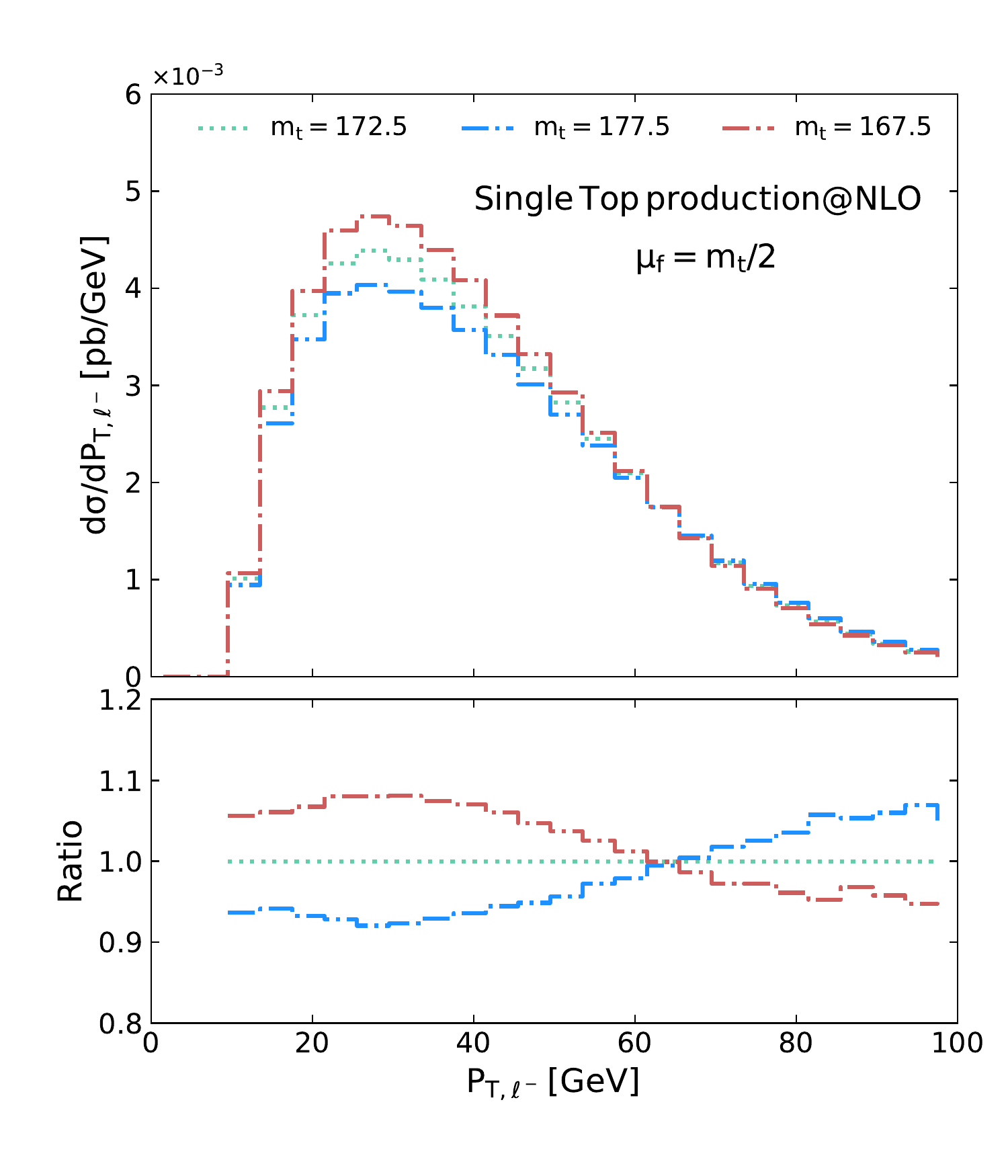}
	\caption{\label{fig:10}  Transverse momentum distribution of the charged lepton with the fiducial	cuts applied for different choices of the top-quark mass at the NLO in QCD.}
\end{figure}

We study the extraction of the top quark mass using the transverse momentum distribution of the charged lepton following our previous work~\cite{Yuan:2020nzd}. We show the sensitivity of the distributions of $P_{T,\ell^-}$ to the top-quark mass in Figure~\ref{fig:10}. The three curves correspond to results of the top-quark mass of 167.5 GeV, 172.5 GeV and 177.5 GeV respectively, with the nominal scale choice. The larger top-quark mass results in a harder  $P_{T,\ell^-}$ distribution and also a lower overall normalization. We choose the average  $P_{T,\ell^-}$ in the fiducial region as our principle observable to extract the top-quark mass. We also study the sensitivity of the total fiducial cross section on the top-quark mass.

The results on average $P_T$  of the charged lepton including their scale variations are presented in Table~\ref{tab:3}. The LO results are less affected by scale choices due to the shape of distribution depends weakly on the scale at the LO. It shows that the LO predictions are not reliable to extract the top-quark mass as can be seen from the gap between the LO and NLO predictions. The NLO scale variations nearly reach the level of changes as induced by varying the top-quark mass by 1 GeV, which is shown in Table~\ref{tab:4}. 

\begin{table}[t]
	\begin{tabular}{|c|c|c|}
		\hline
		[GeV] &LO&NLO\\
		\hline 
		\hline
		$\langle p_{T,\ell^-} \rangle $ & $39.41^{-0.01}_{+0.006}$&$39.75^{-0.11}_{+0.16}$\\
		\hline
	\end{tabular}
	\caption{\label{tab:3}  Average transverse momentum of the charged lepton at various orders in QCD with a central scale choice of $m_t/2$. The superscript (subscript) corresponds to the variation with the scale $m_t$ ($m_t/4$). }
\end{table}

We further investigate the parametric uncertainties of the average $P_{T,\ell^-}$ due to parton distribution functions, bottom-quark mass and  $\alpha_S$ in Table~\ref{tab:4}. The PDF uncertainties are calculated using the CT18NNLO PDFs. In the 5FS, the dependence of the cross section on the bottom-quark mass arises entirely from the bottom-quark PDF which resums the large logarithms of $\alpha_S \ln(Q^2/m_b^2)$ . We study the bottom-quark mass dependence of our predictions using the $\rm MMHT2014NNLO$ PDF set~\cite{Harland-Lang:2014zoa} with a range of $m_b$ values since in CT18 there are no such sets with bottom-quark mass variations. We use three $\rm MMHT$ PDF sets encapsulated in LHAPDF6~\cite{Buckley:2014ana} with bottom-quark masses of 4.25, 4.75 and 5.25 GeV. The resulted variations of the average $P_{T,\ell^-}$ with respect to the one using nominal MMHT PDF set are reported in Table~\ref{tab:4}, and are negligible. We expect the relative variations induced by the bottom-quark mass are not sensitive to the PDF families used since they are mostly of perturbative origin. We calculate the dependence on the QCD coupling constant by varying $\alpha_S(m_Z)$ with 0.118$\pm$0.001. We estimate the statistical error assuming the LHeC can achieve a total integrated luminosity of 100$fb^{-1}$. We conclude that the variations of the average $P_{T,\ell^-}$ by varying the top-quark mass with 1 GeV are significantly higher than the other parametric uncertainties. The statistical error is of similar size as the variation from the top-quark mass. We also show similar results on the total fiducial cross section in Table~\ref{tab:5}. In contrast to the case of average $P_{T,\ell^-}$, the parametric uncertainties are much larger than the variation due to the top-quark mass. It shows about $1\%$ change by varying the top-quark mass with 1 GeV, while the PDF uncertainties are about 2\%. The variations due to the bottom-quark mass are about 5\% if varying $m_b$ by 0.5 GeV, and are 2\% if varying $m_b$ by 0.2 GeV as recommended in Ref.~\cite{Harland-Lang:2015qea}.

\begin{table}
	
	\begin{tabular}{|c|c|}
		\hline
		[GeV]&$\delta \langle p_{T,\ell^-} \rangle$   \\
		\hline 
		\hline
		PDFs(68\% C.L.)&$ + 0.0126 - 0.0081$\\
		\hline		
		$\alpha_S(m_Z)$(0.001) & $+0.0034-0.0031$\\
		\hline
		$m_b$(0.5 GeV)&$\pm$0.0011\\
		\hline
		Statistical error & 0.1341\\
		\hline
		$m_t$(1.0 GeV) & 0.1225\\
		\hline
	\end{tabular}
	
	\caption{\label{tab:4}  Various parametric uncertainties, expected statistical error and dependence on the top-quark mass, of the average $P_{T,\ell^-}$. The parameters are varied by the values in parenthesis. }
\end{table}

\begin{table}
	
	\begin{tabular}{|c|c|}
		\hline
		[\%]&$ \delta\sigma$   \\
		\hline 
		\hline
		PDFs(68\% C.L.)&$ +2.50  -2.08  $\\
		\hline		
		$\alpha_S(m_Z)$(0.001) & $+1.87-1.43 $\\
		\hline
		$m_b$(0.5 GeV)& $\pm$4.86\\
		\hline
		Statistical errors &0.698  \\
		\hline
		$m_t$(1.0 GeV) &0.950 \\
		\hline
	\end{tabular}
	
	\caption{\label{tab:5}  Similar to Table~\ref{tab:4} for the total fiducial cross section. The variations of the cross section are shown in percentages.}
\end{table}

We use a linear model on dependence of the average $P_{T,\ell^-}$ on the top-quark mass

\begin{equation}
	\left\langle p_{T,\ell^-}\right\rangle=p_{T, 0}+\lambda\left[\frac{m_{t}}{\mathrm{GeV}}-172.5\right],
\end{equation}
where $p_{T, 0}$ is the average transverse momentum of the charged lepton for a top-quark mass of 172.5 GeV. The values of $\lambda$ and $p_{T, 0}$ can be obtained from results in Tables~\ref{tab:3} and \ref{tab:4}. The top-quark mass can be extracted from the measurement of the 	$\left\langle p_{T,\ell^-}\right\rangle$. In the following, we estimate the uncertainties of the extracted top-quark mass due to both the statistical error and the theoretical uncertainties. We neglect contributions from various parametric uncertainties which are small as mentioned earlier. 

\begin{figure}[t!]
	\includegraphics[scale=0.47]{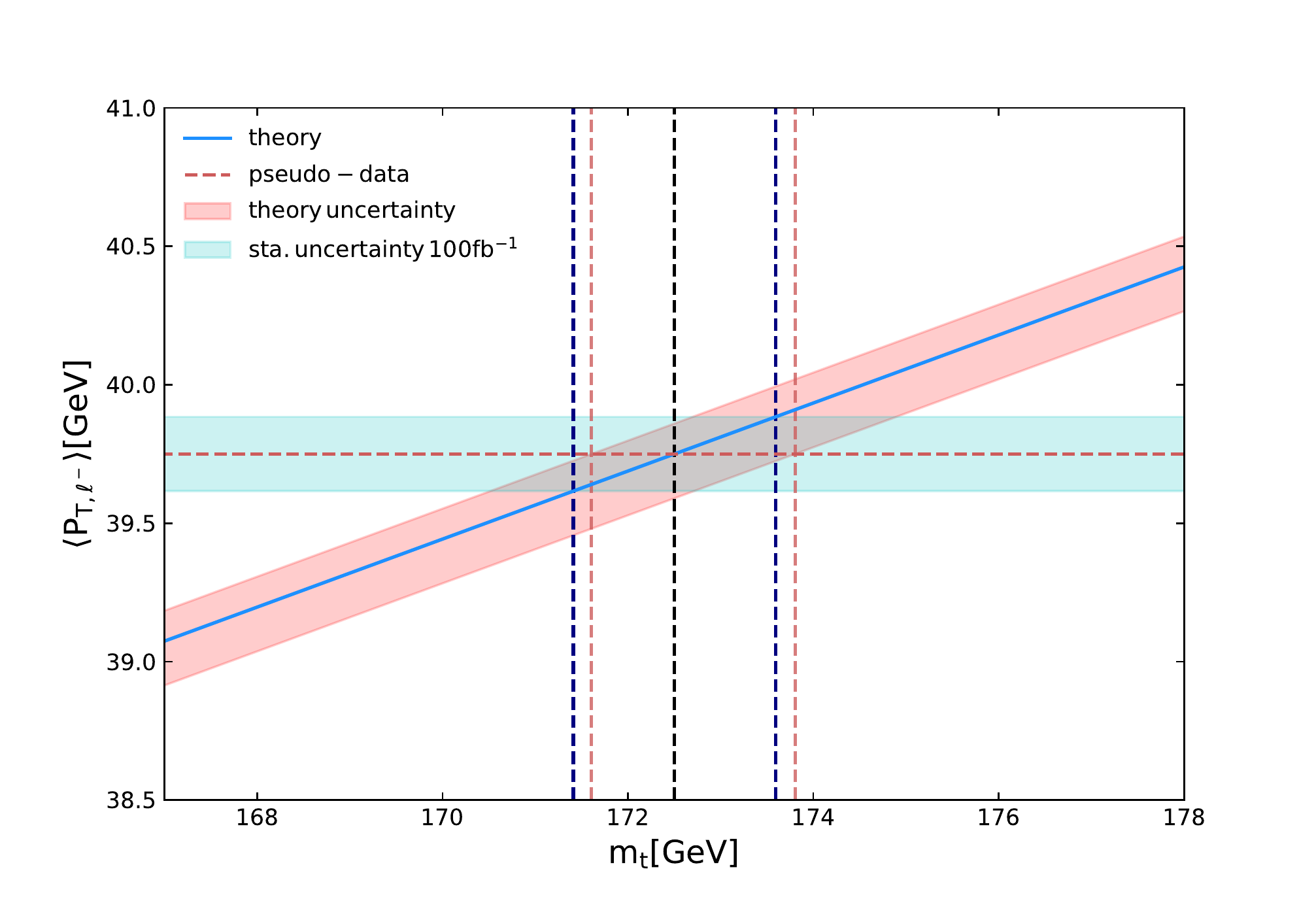}
	\caption{\label{fig:11}   Projection of the top-quark mass measurement with a hypothetical value of 172.5 GeV. The diagonal line is the NLO prediction on $\left\langle p_{T,\ell^-}\right\rangle$ as a function of the top-quark mass. The diagonal band represents the scale variations of  $\left\langle p_{T,\ell^-}\right\rangle$ at the NLO. The horizontal band represents the estimated statistical error of  $\left\langle p_{T,\ell^-}\right\rangle$. Vertical lines indicate various uncertainties of the extracted top-quark mass.}
\end{figure}

We present the projection of the top-quark mass measurement in Figure~\ref{fig:11} with a hypothetical value of  $172.5$ GeV. The horizontal line corresponds to the $\left\langle p_{T,\ell^-}\right\rangle$ for a top-quark mass of 172.5 GeV. The horizontal band represents the estimated statistical error of  $\left\langle p_{T,\ell^-}\right\rangle$. The diagonal blue line is the NLO prediction on $\left\langle p_{T,\ell^-}\right\rangle$  as a function of the top-quark mass. The red band surrounding the diagonal line represents the scale variations of  $\left\langle p_{T,\ell^-}\right\rangle$ at the NLO. The uncertainties on the extracted top-quark mass are computed assuming linear error propagation and are represented by various vertical lines.  The vertical lines of the statistical error are obtained by intersections of the horizontal band and the diagonal line. The vertical lines of the theoretical uncertainty are obtained by intersections of the diagonal band and the horizontal line. Finally, the statistical error of the extracted top-quark mass amounts to 1.1~GeV and the theoretical uncertainty amounts to $+$1.3~GeV and $-$0.9~GeV. The current uncertainties of the direct measurements of the top quark mass at the LHC are about 500-600 MeV~\cite{Zyla:2020zbs,Azzi:2019yne}. The errors of indirect determinations of $m_t$ are about 1-2 GeV~\cite{Zyla:2020zbs}. As for the HL-LHC, the uncertainties of the direct measurements can be reduced to about 200 MeV~\cite{Azzi:2019yne}. These measurements are from top quark pair production. Our proposed determination of $m_t$ is based on the single top quark production with only leptonic observables, and has a precision similar to the LHC indirect measurements.

\subsection{Signal of New Physics}

We study the possible improvement on searches of new physics with the NLO predictions of the single top quark production with leptonic decays. It has been shown that the LHeC can provide a better assessment of the effective $Wtb$ couplings through the measurement of the single top quark production~\cite{Sarmiento-Alvarado:2014eha}. We can write the effective $Wtb$ vertex including SM contributions and those from new physics as~\cite{Bach:2012fb}
\begin{equation}
	\begin{split}
		\mathcal{L}_{t b W}=&-\frac{g}{\sqrt{2}} \bar{b}(\gamma^{\mu}\left((V_{tb}+\delta V_L) P_{L}+V_{R} P_{R}\right)\\
		&-\frac{i \sigma^{\mu \nu} q_{\nu}}{m_{W}}\left(g_{L} P_{L}+g_{R} P_{R}\right)) t W_{\mu}^{-}+h . c . .
	\end{split}
\end{equation}
The new physics contributions can be due to dimension-six effective operators~\cite{Grzadkowski:2010es,AguilarSaavedra:2010zi}, for example~\cite{AguilarSaavedra:2008zc}
\begin{equation}
	\begin{split}
		\frac{1}{\Lambda^{2} } \big\{&C_{\phi q} O_{\phi q}^{(3,33)}+\left[C_{\phi \phi} O_{\phi \phi}^{33}+C_{t w} O_{u W}^{33}\right.\\
		&\left.+C_{b W} O_{d W}^{33}+h . c .\right]\big\},
	\end{split}
\end{equation}
where $\phi$ is the SM Higgs doublet. The definition of the operators can be found in~\cite{AguilarSaavedra:2008zc}. In this case the effective couplings can be related to the corresponding operator coefficients as
\begin{equation}
	\begin{split}
		&\delta V_{L}=\frac{1}{2} \frac{v^{2}}{\Lambda^{2}} C_{\phi q}, \quad g_{R}=\sqrt{2} \frac{v^{2}}{\Lambda^{2}} C_{t W}, \\
		&\quad V_{R}=\frac{1}{2} \frac{v^{2}}{\Lambda^{2}} C_{\phi \phi},\quad g_{L}=\sqrt{2} \frac{v^{2}}{\Lambda^{2}} C_{b W}.
	\end{split}
\end{equation}

It has been shown that the asymmetry of various observables can be sensitive to the modified $Wtb$ couplings, which is defined as~\cite{Sarmiento-Alvarado:2014eha}
\begin{equation}
	A(X,X_0)=\frac{\sigma(X>X_0)-\sigma(X<X_0)}{\sigma(X>X_0)+\sigma(X<X_0)},
\end{equation}
where X is the kinematic observable and $X_0$ is the reference value. We consider the asymmetry of various observables including $\Delta \eta(b,\ell^-)  $, $\Delta \varphi(b,\ell^-)  $, $\Delta \varphi(b,E\!\!\!\!/_T)  $, $\Delta \varphi(\ell^-,E\!\!\!\!/_T)  $ and  cos$(b,\ell^-) $, which are the separation of the pseudorapidities and azimuth angles, and cosine angle of the reconstructed objects  within the fiducial region. We present the SM predictions on the asymmetries $A(\Delta \eta,0)$, $A(\Delta \varphi, \pi/2)$ and $A($cos$,0)$ at both the LO and NLO with scale variations in Table~\ref{tab:6}. The upper(lower) variation corresponds to an alternative scale of $m_t$($m_t/4$). The statistical errors are calculated by assuming a total integrated luminosity of 100$fb^{-1}$. It shows that the NLO corrections are about 7\%-30\% depending on the observables. The scale variations at the LO largely underestimate the perturbative uncertainties. The scale variations at the NLO are about 1\%-3\%, at the same level as the statistical errors.

	\begin{table}
		\begin{tabular}{|c|c|c|c|}
			\hline
			Observable  &LO&NLO&statistical error\\
			\hline 
			\hline
			$\Delta \eta(b,\ell^-)  $ & $-0.374^{-0}_{+0}$&$-0.411^{+0.007}_{-0.008}$&0.006\\
			\hline
			$\Delta \varphi(b,\ell^-)  $ & $0.420^{+0.001}_{-0.002}$&$0.388^{+0.004}_{-0.002}$&0.006\\
			\hline
			$\Delta \varphi(b,E\!\!\!/_T)  $ & $0.805^{-0}_{+0}$&$0.746^{+0.012}_{-0.013}$&0.005\\
			\hline
			$\Delta \varphi(\ell^-,E\!\!\!/_T)  $ & $0.346^{+0}_{-0}$&$0.292^{-0.008}_{+0.010}$&0.007\\
			\hline
			cos$(b,\ell^-)$ & $0.419^{-0.009}_{+0.011}$&$0.548^{-0.018}_{+0.018}$&0.006\\
			\hline
		\end{tabular}
		\caption{\label{tab:6} SM predictions on asymmetries of various observables at both the LO and NLO with a nominal scale choice of $m_t/2$. The numbers correspond to asymmetries $A(\Delta \eta,0)  $, $A(\Delta \varphi, \pi/2)  $ and $A($cos$,0)$, respectively. The upper(lower) variation corresponds to an alternative scale of $m_t$($m_t/4$). The statistical errors are calculated by assuming a total integrated luminosity of 100$fb^{-1}$. }
	\end{table}

We can set constraints on the new physics by comparing the SM predictions with the projected measurements. We calculate the contributions from new physics to the single top quark production with leptonic decays at the LO using MG5\_aMC@NLO~\cite{Alwall:2014hca}. The generated events are analyzed with MadAnalysis5~\cite{Conte:2012fm}. We study the effects of the effective $Wtb$ couplings $\delta V_L$ and $g_R$ using the model SMEFT@NLO~\cite{Degrande:2020evl} assuming the effective couplings are real numbers. The input parameters and fiducial cuts are all set to the same as those in Sec.~\ref*{sec:num}. 

	\begin{table*}
	\begin{tabular}{|c|c|c|c|c|}
		\hline
		&LO&LO + scale variation&NLO&NLO + scale variation\\
		\hline 
		\hline
		$g_R(\sigma_{tot})  $ & [-0.047,0.041]&[-0.236,0.138]&[-0.039,0.036]&[-0.057,0.050]\\
		\hline
		$g_R(\Delta \eta(b,\ell^-) ) $ & [-0.083,0.062]&[-0.083,0.062]&[-0.060,0.051]&[-0.071,0.059] \\
		\hline
		$\delta V_L(\sigma_{tot})  $ & [-0.0086,0.0083]&[-0.034,0.030]&[-0.0073,0.0071]&[-0.010,0.010]\\
		\hline
	\end{tabular}
	\caption{\label{tab:7}  Projected bounds on the effective couplings $g_R$ and $\delta V_L$ at the 95\% CL using different theory predictions by varying the couplings one at a time. The bounds are derived from the measurements on either the total fiducial cross section or the asymmetry $A(\Delta \eta(b,\ell^-),0)$. The statistical errors are included assuming a total integrated luminosity of 100$fb^{-1}$. We further include an experimental uncertainty of 3\% for both the measurement of the fiducial cross section and the asymmetry. }
\end{table*}

We show the projected bounds on the effective couplings $g_R$ and $\delta V_L$ at the 95\% CL using different theory predictions by varying the couplings one at a time in Table~\ref{tab:7}. The bounds are derived from the measurements on either the total fiducial cross section or the asymmetry $A(\Delta \eta(b,\ell^-),0)$. The statistical errors are included assuming a total integrated luminosity of 100$fb^{-1}$. We further include an experimental uncertainty of 3\% for both the measurement of the fiducial cross section and the asymmetry. The theory predictions are based on the new physics contributions calculated at the LO combined with the SM predictions at the LO/NLO and with/without including the scale variations.

We find that the NLO corrections improve the constraints on the new physics in general due to both the shift of the SM central predictions and the reduction of scale variations. At the LO, when including the scale variations, the bounds on $\sigma_{tot}$ are enlarged by almost a factor of four. The NLO predictions result in more reliable bounds. The asymmetry $A(\Delta \eta(b,\ell^-),0)$ yields slightly weaker bounds on $g_R$ comparing to the total fiducial cross section in general. For the bounds on $g_R$ derived from the cross section, the effects of scale variations are larger than in the case of the asymmetry. We have also compared our results with those in~\cite{Sarmiento-Alvarado:2014eha}. Our bounds on $\delta V_L$ and $ g_R$  based on the LO predictions without the scale variations are almost identical to theirs. Note the fiducial cuts used in our study and theirs are slightly different. To summarize we anticipate our best bounds on $g_R$ and $\delta V_L$ at the 95\% CL being [-0.057, 0.050] and [-0.010, 0.010] respectively as derived using the NLO SM predictions with scale variations. The sensitivity of $g_R$ is [-0.05, 0.02] at 95\% CL at the HL-LHC~\cite{Azzi:2019yne}. We conclude that the constraint on the coupling $g_R$ at the LHeC is weaker than the HL-LHC projection. The sensitivity of $\delta V_L$ is [-0.036, 0.036] at the 68\% CL at the HL-LHC as in Ref.~\cite{Sarmiento-Alvarado:2014eha}. We expect the LHeC can give better constraint on $\delta  V_L$ than the HL-LHC.

~\\

\section{Conclusions}
\label{sec:Conclusions}

In this paper, we present a detailed phenomenological study of the single top (anti-)quark production with leptonic decays at the LHeC at NLO in QCD. The NLO calculations are based on the dipole subtraction method and the complex-mass scheme. We include the full off-shell and non-resonant contributions. The NLO corrections reduce the inclusive cross section by 8.5\%. While in a typical fiducial region, the NLO corrections reduce the cross section by 14\%. We also present predictions of various distributions. The NLO predictions exhibits strong stability under scale variations for both the total cross section and the distributions. The PDF uncertainty can be larger than the scale variations at the NLO depending on the kinematic region considered. 

Moreover, we study the extraction of  the top-quark mass from  measurement of the average transverse momentum of the charged lepton. We find that the statistical error of the extracted top-quark mass amounts to 1.1 GeV. The theoretical uncertainty due to the scale variations at the NLO are +1.3 GeV and $-$0.9 GeV. The uncertainties due to input parameters including the $\alpha_S$, bottom-quark mass and PDFs are all negligible. Besides, we study the possible improvement on searches of new physics with the NLO predictions of the single top quark production with leptonic decays. We obtain better constraints on $Wtb$ effective couplings $g_R$ and $\delta V_L$ using the NLO predictions comparing with the LO predictions. The bounds derived using the LO predictions are much weak when including the scale variations. We anticipate our best bounds on $g_R$ and $\delta V_L$ at the 95\% CL being [-0.057, 0.050] and [-0.010, 0.010] respectively as derived using the NLO SM predictions with scale variations. 

\acknowledgments

The work of JG is sponsored by the National Natural Science Foundation of China under the Grant No. 11875189 and No. 11835005. MG acknowledges valuable conversations with Chuanle Sun, Shurun Yuan and Shuo Lou.

\bibliography{references}
\end{document}